# A Study on Privacy-Preserving Scholarship Evaluation Based on Decentralized Identity and Zero-Knowledge Proofs

Yi Chen
College of Electronics
and Information Engineering
Shenzhen University
Shenzhen, China
chenyi_0916@163.com

Bin Chen *
College of Electronics
and Information Engineering
Shenzhen University
Shenzhen, China
bchen@szu.edu.cn
*Corresponding author

Peichang Zhang
College of Electronics
and Information Engineering
Shenzhen University
Shenzhen, China
pzhang@szu.edu.cn

Da Che
College of Art and Design
Shenzhen University
Shenzhen, China
cheda@szu.edu.cn

*Abstract*—Traditional centralized scholarship evaluation processes typically require students to submit detailed academic records and qualification information, which exposes them to risks of data leakage and misuse, making it difficult to simultaneously ensure privacy protection and transparent auditability. To address these challenges, this paper proposes a scholarship evaluation system based on Decentralized Identity (DID) and Zero-Knowledge Proofs (ZKP). The system aggregates multi-dimensional ZKPs off-chain, and smart contracts verify compliance with evaluation criteria without revealing raw scores or computational details. Experimental results demonstrate that the proposed solution not only automates the evaluation efficiently but also maximally preserves student privacy and data integrity, offering a practical and trustworthy technical paradigm for higher education scholarship programs.

## I. INTRODUCTION

As data privacy receives greater attention, users expect better protection for their personal information. Yet most existing scholarship review systems still rely on a centralized architecture, requiring students to submit sensitive information such as identity details, academic performance, research results, and social practice records to the academic affairs office or various functional departments for unified storage and review [1]. This approach improves efficiency but risks data leakage and misuse. To enhance transparency and security, platforms like ScholarChain [2] and CryptoScholarChain [3] have migrated these workflows on-chain by using smart contracts to automate application, verification and disbursement. However, since some application information or proof information still needs to be made public on the chain, there is room for improvement in the protection of students' private data.

Decentralized Identity (DID) [4] gives students full control over their personal data by allowing them to decide exactly which attributes such as student number, enrollment status or department affiliation are shared, with whom and for how long. Departments, in turn, issue Verifiable Credentials (VCs) that contain only the necessary metadata for example, "student X has a GPA credential issued on date Y" without including the actual numeric score [5][6]. This selective disclosure model minimizes data exposure but does not prevent a verifier from demanding raw scores.

Zero-Knowledge Proofs (ZKP) fill this gap by enabling students to cryptographically prove claims about their data such as "GPA ⩾ 3.5" without revealing the exact values [7][8]. In practice, a student constructs a ZKP using their private score and a public threshold, then submits only the succinct proof and public parameters to the smart contract. The contract runs a verification function that accepts or rejects the proof in a single transaction, automatically enforcing eligibility rules without ever learning the student's precise grades.

This paper presents a scholarship evaluation system that combines DID and ZKP. In this system, students manage their identities through DID and generate ZKPs that hide individual dimension scores while validating weighted totals. This design protects privacy and supports an automated and auditable process. The remainder of the paper is organized as follows. Section II describes the system architecture and ZKP circuit design; Section III defines the security model; Section IV discusses deployment and performance evaluation; Section V concludes.

## II. SYSTEM MODEL

To enable a decentralized, privacy-preserving scholarship evaluation process, the proposed system integrates DID and ZKP technologies into a unified framework. Fig. 1 depicts the overall workflow of the proposed scholarship evaluation system, involving three principal roles: Credential Authorities (CAs), the Administrator, and Students. CAs are trusted third-parties entities that issue Verifiable Credentials (VCs) and generate dimension-specific Zero-Knowledge Proofs (ZKPs) upon student authorization. The Administrator is responsible for deploying the scholarship smart contract and issuing final scholarship VCs to eligible students. Students, as the core participants, authorize CAs, generate the aggregation proof, submit applications, and claim their scholarship credentials.

The proposed system assumes that each student's total scholarship score is computed as a weighted sum across multiple dimensions, including but not limited to course grade, research achievements, volunteer service, and competition performance. Each dimension's raw score is evaluated by the corresponding CA according to predefined criteria and issued as a VC attesting to its validity. Before applying, students complete KYC and register a campus DID on-chain, which they use to sign and verify all subsequent actions.

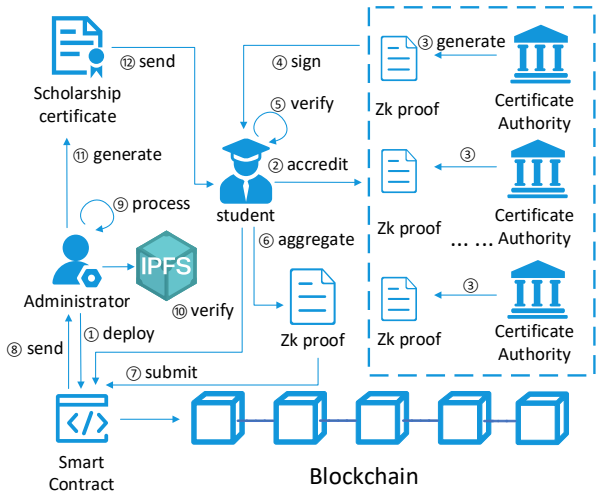

Fig. 1. The processes of the scholarship evaluation system.

The system comprises the following five core phases:

*A. Smart Contract Deployment*

In order to automate and decentralize the scholarship evaluation process, the Administrator first deploys the scholarship smart contract on the blockchain. The main contract manages the end-to-end workflow, including parameter configuration, application control, proof submission and verification, result selection, and credential issuance. Auxiliary contracts handle the verification and management of each CA's public key and ZKP verification key, as well as the storage and validation of tiered Merkle roots and IPFS references.

During deployment, the main contract must initialize critical parameters. TABLE I. lists the primary parameters involved in the main contract deployment along with their descriptions.

TABLE I.    PARAMETER OF SMART CONTRACT

| Parameter | Detail |
|---|---|
| ID | Unique identifier of the Scholarship |
| Applicant | Store the DID and score of the applicant |
| PrizeCount | Store the number of winners for each award tier |
| AwardRoots | Store Merkle tree roots of final winner |
| Weight | Store the weight of scores in each dimension |
| CAPubKey | Store the public keys of each Certificate Authority |
| StartTime | Start time of scholarship application |
| EndTime | End time of scholarship application |

*B. Authorization for Proof Generation*

The specific process is illustrated in Fig. 2. Students use their DID, identity VC, and corresponding VC IDs to authorize each CA to generate zero-knowledge proofs. Taking the $i$-th CA as an example, let its private key be $sk_i$ and its public key be $pk_i$.

After completing identity verification, the CA locates the

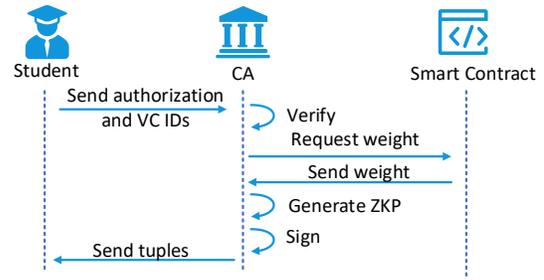

Fig. 2. The ZKP generation processes of CA.

corresponding VC by its ID and extracts the score $s_i$ from the VC. It retrieves the weight $w_i$ from the scholarship smart contract. The CA executes the pre-defined zero-knowledge circuit according to the following algorithm:

**ALGORITHM 1**: Weighted Score Algorithm

**Private Input:** The score $s_i$
**Public Input:** The weight $w_i$
**Output:** The weighted score $s_{w_i}$
start
    set $s_{wi} = 0$
    verify $0 \leq s_i \leq 100$
    verify $0 \leq w_i \leq 100$
    $s_{w_i} = s_i \times w_i$
    verify $0 \leq s_{w_i} \leq maxScore$
    return $s_{wi}$
end

The CA executes the circuit to generate a ZKP $\pi_i$, demonstrating that the score for the corresponding dimension satisfies the scholarship weighting rule:

$$\pi_i = Prove(\, pk_{cicuit}\, ,s_i\, ,w_i\, ,s_{w_i}) \quad (1)$$

$s_{wi}$ represents the weighted score. The original score $s_i$ is a private input, while $w_i$ and $s_{wi}$ serve as public inputs for $\pi_i$.

The proof $\pi_i$ is concatenated with its encoded public inputs $Encode(w_i\, ,s_{w_i})$, and a hash $h_i$ is computed as follows:

$$h_i = Hash\left(\pi_i \,\|\, Encode(w_i\, ,s_{w_i})\right) \quad (2)$$

The CA sign $h_i$, producing the signature $\sigma_i$. Then the CA packages the tuple $T_i = (\pi_i, w_i, s_{w_i}, h_i, \sigma_i, pk_i)$ and returns it to the student.

*C. Proof Aggregation & Submission*

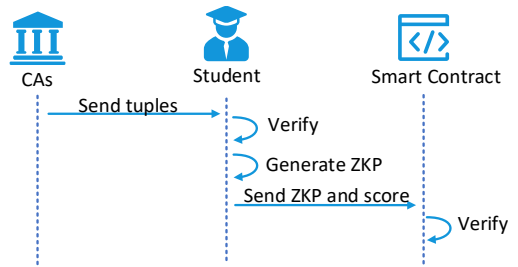

Fig. 3. The ZKP aggregation processes of student.

Fig. 3 shows the process. Upon receiving different tuple

$T_i$ from different CA, the student recomputes each hash $h'_i$ and verifies each signature $\sigma_i$. If the following conditions hold:

$$Verify(pk_i, h_i, \sigma_i) = 1 \tag{3}$$

then the student computes the final total score as:

$$s_{total} = \sum_{i=1}^{n} s_{w_i} \tag{4}$$

Next, these inputs are fed into the aggregation circuit, where the detailed computation is performed as follows:

---
**ALGORITHM 2**: Aggregate Algorithm
---
**Private Input:** the proofs $\{\pi_1, \pi_2, ..., \pi_n\}$, the weight values set by the scholarship $\{w_1, w_2, ..., w_n\}$, the score after weighting $\{s_{w_1}, s_{w_2}, ..., s_{w_n}\}$ and the signatures of CAs $\{\sigma_1, \sigma_2, ..., \sigma_n\}$
**Public Input:** CAs' public key $\{pk_1, pk_2, ..., pk_n\}$, hash values $\{h_1, h_2, ..., h_n\}$, and total score $s_{total}$
start
   $i = 1$, $s_{temp} = 0$
   while( i ⩽ n )
      $h_{temp} = Hash(\pi_i \parallel Encode(w_i, s_{w_i}))$
      if ( $h_{temp}$ equals $h_i$ and $Verify(pk_i, h_i, \sigma_i) = 1$)
         $s_{temp} = s_{temp} + s_{w_i}$
      i=i+1
   if ( $s_{temp}$ equals $s_{total}$)
      return $s_{total}$
   else
      return false
end

---

The student executes the aggregation circuit to generate a zero-knowledge proof $\pi_{agg}$:

$$\pi_{agg} = Prove(pk_{agg}, \{\pi_i, w_i, s_{w_i}, \sigma_i\}_{i=1}^{n}, \{pk_i, h_i\}_{i=1}^{n}, s_{total}) \tag{5}$$

Where $pk_{agg}$ is the proving key for the aggregation circuit.

The student submits the proof $\pi_{agg}$, the public inputs $\{pk_i, h_i\}_{i=1}^{n}$, and the total score $s_{total}$ to the smart contract. A single on-chain verification ensures that all proofs originate from trusted CAs and remain unaltered, and that their sum matches $s_{total}$. The entire process is depicted in Fig. 3.

### D. Result Selection

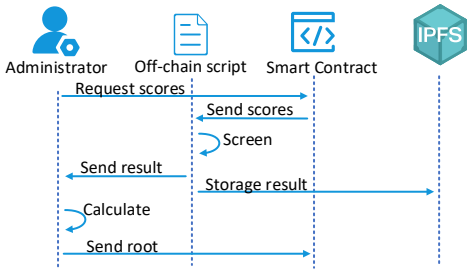

Fig. 4. The processes of Result Selection.

This process is illustrated in Fig. 4. After the application deadline, the Administrator retrieves all application records including each student's DID and their $s_{total}$ from smart contract and packages this data for off-chain processing. The off-chain script receives the list, sorts it in descending order by $s_{total}$, and allocates students to different award tiers according to predefined quotas. The finalized awardee lists are then uploaded to IPFS [9].

For each award tier $L$ (with $M_L$ awardees), the system computes a leaf node for each awardee as:

$$leaf_i = Hash(student\ DID \parallel s_{total i}) \tag{6}$$

A Merkle tree is then constructed over all leaf nodes $\{leaf_1, leaf_2, ..., leaf_{M_L}\}$ to obtain the Merkle root $root_L$[10]. The Administrator uploads $root_L$ to the smart contract.

### E. Credential Issuance

The awarded students can query the full list of recipients and their corresponding leaf nodes $\{leaf_1, leaf_2, ..., leaf_{M_L}\}$ via the system frontend or through IPFS. Each student computes their own leaf node $leaf_i$ using their DID and total score $s_{total}$, and generates the corresponding Merkle proof $proof_i$. They then submit both $leaf_i$ and $proof_i$ to the smart contract for verification. Upon successful verification, the student receives the scholarship VC issued by the Administrator.

## III. SECURITY ANALYSIS

In this section, the security of the decentralized scholarship evaluation system is analyzed from four perspectives:

*1) Unforgeability:* In our system, CAs are fully trusted and securely hold their private key. Any forge signature or proof fails ZKP verification, hash checks, or signature validation on-chain. Since ZKPs and digital signatures rely on hard cryptographic assumptions, forging them is computationally infeasible, ensuring proof and signature unforgeability.

*2) Privacy Preservation:* All dimension scores $s_i$ are verified via ZKPs. The details of each dimension's computation remain entirely within the circuit. On-chain, only the total score $s_{total}$ and the Merkle root $root_L$ are published; no individual sub-scores are stored or exposed.

*3) Program Correctness:* Smart contracts automate application, verification, and tier assignment under immutable, publicly inspectable rules. While the Administrator triggers sorting, any observer can replicate the results using the same algorithm.

*4) Verifiability:* Smart contracts and off-chain scripts ensure full auditability. Events such as application submissions, Merkle root updates, and credential claims are publicly visible. Anyone can fetch the awardee list, compare the Merkle root, and verify consistency. The open-sourced circuit code and Verifier Key also enable independent third-party audits of proof aggregation.

## IV. EXPERIMENTAL ANALYSIS

We deploy our system on the FISCO BCOS consortium blockchain, implementing scholarship evaluation, credential verification, and disbursement in Solidity. Decentralized identity is managed via the WeIdentity framework [11]. Blockchain nodes run on Ubuntu 18.04 (Intel Xeon Platinum

8269, 4 GB RAM), while clients use Windows 10 (Intel Core 3.00 GHz, 48 GB RAM). For ZKPs, we use Zokrates 0.8.8 todesign circuits and generate ZK-SNARK proofs.

### A. Functionality Verification

This part illustrates the scholarship application workflow using two students, Alice and Bob, who complete the process from application to credential issuance via the proposed platform. Both are assumed to have completed KYC and registered their DIDs on the blockchain. TABLE II. lists the DIDs for Alice, Bob, and the other primary roles in the system.

TABLE II. ROLES AND THEIR CORRESPONDING DIDS

| Role | DID |
|---|---|
| Alice | did:weid:666:0x9dae21d...7d5b8477a81b1c626f00f0 |
| Bob | did:weid:666:0x53622ca...b910951ed368cbd6d606a8 |
| Administrator | did:weid:666:0x8770a36...dac4bcd3f420276bbfb8f5d |
| $CA_1$ | did:weid:666:0x890897b...12aab1c5f2449d55c992ee |
| $CA_2$ | did:weid:666:0x22db649...cbc6c20990b45fb0771b81 |
| $CA_3$ | did:weid:666:0xa687c92...56fe8a6e097a533a4d7647 |
| $CA_4$ | did:weid:666:0x36fcebb...784c318a42744f7f0858fd8 |

Next, TABLE III. defines the four scholarship review dimensions and their corresponding weights:

TABLE III. SCHOLARSHIP REVIEW DIMENSIONS AND WEIGHTS

| Dimensions | Weights $w_i$ |
|---|---|
| Percentage Score $w_1$ | 60 |
| Research Output $w_2$ | 20 |
| Volunteer Service $w_3$ | 10 |
| Competition performance $w_4$ | 10 |

TABLE IV. shows sample raw scores for Alice and Bob, serving as inputs for the CAs when they generate the proofs:

TABLE IV. SAMPLE RAW SCORES

| Applicant | Percentage Score $s_1$ | Research $s_2$ | Volunteer $s_3$ | Competition $s_4$ |
|---|---|---|---|---|
| Alice | 95 | 85 | 40 | 92 |
| Bob | 80 | 90 | 60 | 90 |

After these preparations, each CA generates its ZKP based on the scores in TABLE IV. and the weights in TABLE III. To illustrate, we take Alice's Percentage Score as an example. Here, Alice's raw score is $s_1 = 95$ and the scholarship weight is $w_1 = 60$. $CA_1$ uses these two values computes the weighted score $s_{w_i}$ and produces the proof in Fig. 5.

```
{
"scheme": "g16",
"curve": "bn128",
"proof": {
  "a": [ "0x10d76181ad45ec5ab14d4c44ce70052819548f7467b733ec161b82f20da6496a",
         "0x21920bdc48dacd9e721c434cd4f89cd530d3b4218e89e804556bc5512650e7f4"],
  "b": [["0x0594efad85c51ec82a2695d8003e3beb9edc911119aa3d931b0b15d4a70db4ca",
         "0x18e0ed7dfc38e6797e6b74c98f1f7ca56abbcc5b2663c164f6a7577e80d1c48d"],
        ["0x014f1e2c56014e8490d109b38efa279de7831a8014c22abfc317589f69d9e367",
         "0x07aa6df3a221284474b3af6fa17329838 1abeef23930836a44986f35d485db6e"]],
  "c": [ "0x140cff39419d6412acff943d1495bfb2db578eb682d5b59704017d68ddc9e8a3",
         "0x3013c6a3112a60349555600e4a9321feb9b8465f6ca79f906acc11eda904f962"]}
}
```

Fig. 5. The ZKP of the weighted score.

$CA_1$ concatenates the proof $\pi_1$ with $(w_i, s_{wi})$, computes the hash $h_1$ and signs $h_1$. It then returns the tuple $T_1$ to Alice. The same procedure is carried out by each corresponding CA for the other dimensions.

Once Alice has collected all credential packages, she verifies signatures and recomputes hashes. Upon successful checks, she submits the proofs and public inputs $(w_i, s_{wi})$ to the aggregation circuit, which validates them in one go and outputs the aggregate proof $\pi_{agg}$ (in Fig. 6). For Alice, the circuit returns is $s_{total1} = 8720$, which represents 87.20 after accounting for the circuit's integer-only arithmetic.

```
{
"scheme": "g16",
"curve": "bn128",
"proof": {
  "a": [ "0x04aa6658b71be5d810f1232ff26e56e37ee9a9f4c342d6a8d46266605b261d31",
         "0x0b75a78676ccd5fc2049f55917058c741fc251a68ef209d851528016d880f1f8"],
  "b": [["0x1e2f4e0650a7351fd92541637e154a546c5fa2ad2f3710acba579b5b17004756",
         "0x153fbbe553c0f86f100860a97b56944de9c190313ce6e5f7da587bd0c58188c8"],
        ["0x2dda8628c2012f4fa2de0cb9a839722c81ce4dd55acfb733e627e2a9a372ce22",
         "0x08fc66384a93101d4de6aa9858e1bdad540d7b814682ab40f8820f5d36b414cb"]],
  "c": [ "0x015ffa068417591b48239913425d198f83145e6c20c73c148d8b6852e7866fcf",
         "0x2a3e7b154550d60b8c22d4fe6aba5c0c16955094d811c1111d5237e620753ec5"]}
}
```

Fig. 6. The ZKP of the Aggregation Verification.

Alice packages the proof $\pi_{agg}$ together with the public inputs $\{pk_i, h_i\}_{i=1}^{4}$ and her total score $s_{total1}$, then submits them to the on-chain contract to apply for the scholarship. Bob follows the same steps with $s_{total2} = 81.0$.

The smart contract performs an on-chain verification of $\pi_{agg}$. Upon successful validation, it records each applicant's DID and submitted total score for subsequent result selection. After the application deadline, the contract automatically retrieves all verified applications and ranks them in descending order of score. Alice receives the First-Class Scholarship and Bob the Second-Class Scholarship. An off-chain script uploads the awardee list to IPFS and records its Merkle root on-chain. Alice and Bob each retrieve their Merkle proofs, submit them for on-chain verification, and upon validation receive their scholarship VCs. Fig. 7 shows the scholarship VC received by Alice; Bob's credential follows the same format.

```
{"claim": {
    "applyScore": 87.2,
    "claimTime": "2025-06-26",
    "level": "firstPrize",
    "scholarshipID": 5,
    "scholarshipName": "Academic Scholarship",
    "studentDID": "did:weid:666:0x9dae21d046efa801bd7d5b8477a81b1c626f00f0"
  },
  "context": "https://github.com/WeBankFinTech/WeIdentity/blob/master/context/v1",
  "cptId": 1014,
  "expirationDate": "2029-06-26T16:02:50Z",
  "id": "3cb24d86-52e0-4b9a-a32b-fbaff308bbae",
  "issuanceDate": "2025-06-26T16:02:50Z",
  "issuer": "did:weid:666:0x8770a3604cb6cbee1dac4bcd3f420276bbfb8f5d",
  "proof": {
    "created": "2025-06-26T16:02:50Z",
    "creator": "did:weid:666:0x8770a3604cb6cbee1dac4bcd3f420276bbfb8f5d",
    "salt": {
      "applyScore": "L9G6z",
      "claimTime": "VLXV8",
      "level": "xVH4I",
      "scholarshipID": "IfdAT",
      "scholarshipName": "ptnZ4",
      "studentDID": "OBv2C"
    },
    "signatureValue": "YZlMEGFNtIO5ga4PfD/CgWUmgCkiyH/
      ZeMc5Jvq8Lo4IsFIsjnvokBad/nOCahGD6L15Jn4kJ3wa8Vqx0/kYgQA=",
    "type": "Secp256k1"
  },
  "type": ["VerifiableCredential", "original"],
  "$from": "toJson"
}
```

Fig. 7. The scholarship VC for Alice.

## B. Performance Analysis

To provide a clear performance overview, we divide our evaluation into two parts. The on-chain analysis examines the smart contract's performance, while the off-chain analysis focuses on the ZKP workflow.

### 1) On-Chain Performance

TABLE V. compares the processing time of the scholarship smart contract under different levels of concurrent application submissions. It takes 56 ms for a single request, rising to 294 ms (10), 360 ms (30), 567 ms (60), and 804 ms (100). This increase stems from handling parallel requests: more transactions to order, signatures to verify, state updates to process, and I/O to carry out, all of which add latency.

TABLE V. THE PERFORMANCE OF SMART CONTRACT

| Current user number | 1 | 10 | 30 | 60 | 100 |
|---|---|---|---|---|---|
| Time(ms) | 56 | 294 | 360 | 567 | 804 |

To evaluate the on-chain cost of credential issuance, we measured the gas consumption of the claimScholarship function. A single transaction consumed 50,642 gas, with 28,530 gas used for function execution. For comparison, a simple ERC-721 minting on the same network uses approximately 56,000 gas. These data indicate that the computation and costs required for credential issuance remain within acceptable bounds.

### 2) Off-Chain Performance

To assess the off-chain performance, we benchmark two ZK circuits: the basic weighted-score circuit and the aggregation circuit.

TABLE VI. THE PERFORMANCE OF THE WEIGHT CIRCUIT

|  | Generate proofs(ms) | Verify proofs(ms) |
|---|---|---|
| Time | 82 | 9 |

TABLE VI. illustrates the average proof-generation and verification times for the weighted-score circuit over multiple runs. On average, generating a single proof takes approximately 82ms, while verification requires only about 9ms. The lightweight nature of the verification step implies that a large number of proofs can be processed in parallel off-chain without becoming a performance bottleneck.

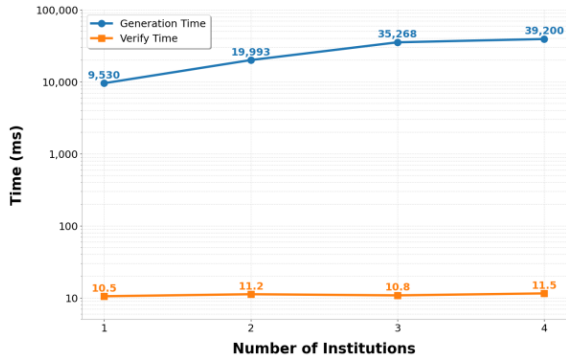

Fig. 8. The performance of the aggregation circuit

Fig. 8 shows proof generation and verification times as the number of CAs increases from 1 to 4. Generation time rises sharply from 9.5 s (1 CA) to 19.9 s (2 CAs), 35.2 s (3 CAs), and 39.2 s (4 CAs)—because each extra CA adds costly hash, signature, and accumulation steps, expanding circuit constraints and large integer work. Verification time, however, stays steady at 10–12 ms, showing the Verifier Key's one time check incurs minimal, constant overhead.

## V. CONCLUSION

This paper has presented the design and implementation of a privacy-preserving scholarship evaluation system that integrates Decentralized Identity and Zero-Knowledge Proofs. By leveraging DID, the system enables students to self-manage their identity data in a secure and flexible manner, while ZKP allows applicants to prove compliance with scholarship criteria without disclosing any sensitive personal information.

## ACKNOWLEDGMENT

This work is supported in part by the Foundation of Guangdong Province Graduate Education Innovation Program Project[2024JGXM_163], Shenzhen under Grant 20220810142731001, Shenzhen University Graduate Education Reform Research Project [SZUGS2023JG02]. Shenzhen University High-Level University Construction Phase III -Human and Social Sciences Team Project for Enhancing Youth Innovation[24QNCG06].